\documentclass[
preprint,
showpacs,preprintnumbers,
bibnotes,
 amsmath,amssymb,
 aps,
 pra,
superscriptaddress,
longbibliography,
]{revtex4-1}

\usepackage{amsmath}
\usepackage{amsfonts}
\usepackage{amssymb}
\usepackage{graphicx}
\usepackage{xspace}
\usepackage{color}
\usepackage{comment}
\usepackage[hidelinks]{hyperref}
\usepackage{dcolumn}
\usepackage{upgreek}
\usepackage{bm}
\usepackage{mathtools}

\usepackage{comment}

\usepackage[detect-weight=true, detect-family=true]{siunitx}

\graphicspath{ {./figures/} }

\begin{document}

\title{Tip-enhanced quantum-sensing spectroscopy for bright and reconfigurable solid-state single-photon emitters}
\author{Hyeongwoo Lee$^{\dagger}$}
\affiliation
{Department of Physics, Pohang University of Science and Technology (POSTECH), Pohang 37673, Republic of Korea}
\author{Taeyoung Moon$^{\dagger}$}
\affiliation
{Department of Physics, Pohang University of Science and Technology (POSTECH), Pohang 37673, Republic of Korea}
\author{Hyeonmin Oh}
\affiliation
{Department of Physics, Pohang University of Science and Technology (POSTECH), Pohang 37673, Republic of Korea}
\author{Kijeong Park}
\affiliation
{Department of Physics, Pohang University of Science and Technology (POSTECH), Pohang 37673, Republic of Korea}
\author{Huitae Joo}
\affiliation
{Department of Physics, Pohang University of Science and Technology (POSTECH), Pohang 37673, Republic of Korea}
\author{Milos Toth}
\affiliation
{School of Mathematical and Physical Sciences, Faculty of Science, University of Technology Sydney, Ultimo, New South Wales 2007, Australia}
\affiliation
{ARC Centre of Excellence for Transformative Meta-Optical Systems (TMOS), University of Technology Sydney, Ultimo, New South Wales 2007, Australia}
\author{Igor Aharonovich}
\affiliation
{School of Mathematical and Physical Sciences, Faculty of Science, University of Technology Sydney, Ultimo, New South Wales 2007, Australia}
\affiliation
{ARC Centre of Excellence for Transformative Meta-Optical Systems (TMOS), University of Technology Sydney, Ultimo, New South Wales 2007, Australia}
\author{Kyoung-Duck Park$^{\ast}$}
\affiliation
{Department of Physics, Pohang University of Science and Technology (POSTECH), Pohang 37673, Republic of Korea}
\affiliation
{Department of Semiconductor Engineering, Pohang University of Science and Technology (POSTECH), Pohang 37673, Republic of Korea}
\affiliation
{Center for Multidimensional Carbon Materials (CMCM), Institute for Basic Science (IBS), Ulsan 44919, Republic of Korea}
\affiliation
{Institute for Convergence Research and Education in Advanced Technology, Yonsei University, Seoul 03722, Republic of Korea}

\begin{abstract}

\noindent 
\textbf{
Atom-like defects in hexagonal boron nitride (hBN) provide room-temperature single-photon emission and coherent spin states, making them attractive for quantum-computing and -sensing applications. 
However, their random spatial and spectral characteristics hamper deterministic coupling with nano-optical cavities, limiting their use as bright single-photon sources and sensitive quantum sensors.
Here, we present tip-enhanced quantum-sensing spectroscopy of single-photon emitters in hBN.
Through precise spatial positioning of individual emitters within tip-cavities with different plasmon resonances, we adaptively control the enhancement rates of both excitation and emission, as well as the single-photon purity.
In this way, optimal selection of their relative contributions can effectively reconfigure solid-state single-photon sources, with simultaneous nano-spectroscopic space- and time-resolved analyses.
Furthermore, we demonstrate tip-enhanced quantum-sensing with single spin defects through optically detected magnetic resonance (ODMR) experiments in tip-coupled hBN nanoflakes.
Our approach provides a unique pathway toward highly-sensitive and deterministic quantum-sensing with room-temperature single-photon emitters.
}

\end{abstract}

\maketitle

\noindent 
Solid-state single-photon emitters are the essential building blocks of quantum photonic technologies, e.g., quantum dots serving as a prime example for a range of quantum applications \cite{o2009, garcia2021, loss1998, imamog1999, lodahl2015}. 
While these artificial atoms offer bright, coherent single-photon emission for quantum communication and computing \cite{loss1998, stevenson2006}, addressing the requirement for cryogenic operation and suppressing spectral broadening and fluctuation remain essential for realizing scalable and practical implementations \cite{borri2001, thoma2016}.
Recently, color centers in wide-bandgap semiconductors have emerged as a compelling alternative to conventional quantum emitters, owing to their stable optical and spin properties that persist even under ambient conditions.
These atom-like defects can exhibit narrow zero-phonon lines (ZPLs) and long spin coherence, realizing practical room-temperature applications in quantum information processing and sensing \cite{bar2013solid, mizuochi2012, pezzagna2011, aharonovich2016}.
Notably, color centers hosted in van der Waals (vdW) materials, particularly atomic defects in hexagonal boron nitride (hBN), offer additional advantages derived from the 2-dimensional host lattice, including outstanding scalability for integration with nano-photonic and -plasmonic structures \cite{tran2016, montblanch2023}.
Moreover, they provide ideal platforms for exploring cavity quantum electrodynamics (cQED) at room temperature, laying the groundwork for next-generation quantum computing and sensing \cite{blais2021, blais2020, wang2024}.

Plasmonic nanocavities represent a particularly appropriate platform for harnessing these advantages.
Localized surface plasmon resonance (LSPR) confines optical modes to sub‐wavelength volumes, producing intense near‐field enhancement and elevated local density of optical states, which increase the excitation and spontaneous emission rates, respectively.
Furthermore, the intrinsic dissipative losses in plasmonic cavities result in broad resonance linewidths, thus relaxing detuning constraints encountered in high-Q dielectric systems.
This broad spectral tolerance facilitates the adaptive optimization of excitation and/or emission rates through spectral matching with the plasmon resonance \cite{akselrod2014, russell2012, li2015}.
Consequently, plasmonic systems cavities can modify key performance metrics of quantum photonic technologies, such as brightness, purity, and coherence \cite{akselrod2014, hoang2015, bogdanov2019}.
Despite these promising attributes, conventional static plasmonic nanocavities face significant challenges when interfaced with atom-like defects in hBN.
Coupling quantum emitters to plasmonic nanocavities requires a more rigorous approach than classical light-emitting systems, given the extreme sensitivity of quantum characteristics to external perturbations \cite{hennessy2007, mendelson2022, wolf2015}. 
For example, achieving bright, high-purity single-photon emission demands precisely engineered photonic environments that optimally balance excitation and emission enhancements. 
Population shelving into metastable states increases nonradiative decay, leading to photon bunching and a reduced antibunching contrast, thereby lowering the single-photon purity \cite{tran2016, maragkou2012, babinec2010, kurtsiefer2000}.
However, the inherently random formation of defects in hBN leads to considerable spatial and spectral variations \cite{tran2016, stern2019, tran2016robust}, posing a major challenge.  
Overcoming these barriers is crucial to fully harness the potential of plasmonic systems and bridge the gap between the fundamental research and the practical realization of room-temperature quantum technologies.

Here, we present tip-enhanced quantum-sensing spectroscopy that deterministically couples a tip-cavity with atomic defects in hBN and enables real-time spectroscopic sensing of their cavity-induced single-photon emission properties.
This approach combines 3-dimensional spatial positioning of the emitter within tip-cavities with different plasmon resonances, allowing selective enhancement of excitation and/or emission rates.
By tuning the spectral overlap among the emitter transition, the excitation laser, and the plasmon resonance, we select coupling regimes in which either radiative decay or optical pumping dominates. 
In the emission-dominated regime, Purcell-accelerated decay shortens the excited-state lifetime and suppresses photon bunching, thereby improving single-photon purity. 
In the excitation-dominated regime, near-field pumping promotes bunching and reduces purity.
Furthermore, we present a simple quantum-sensing platform by attaching hBN nanoflakes to the apex of an Au tip and performing optically detected magnetic resonance (ODMR) measurements, achieving a magnetic-field sensitivity of $\sim$116 $\mu$\textit{T}$/$$\sqrt{\textnormal{Hz}}$.
These approaches propose a versatile platform for dynamically controlling solid-state quantum emitters and precisely sensing their properties for advanced quantum technologies.
\\

\begin{figure*}[!t]
	\includegraphics[width = 16.5 cm]{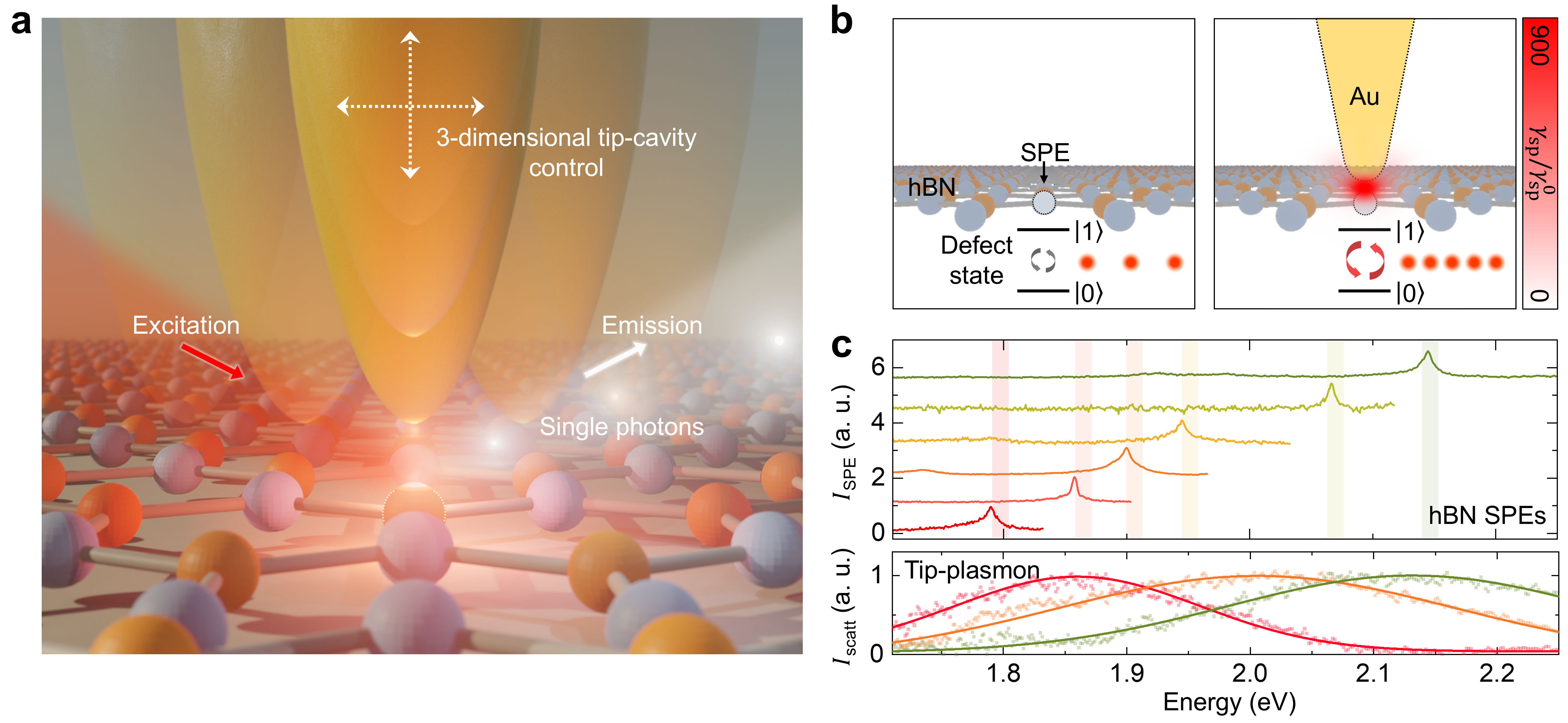}
	\caption{
\textbf{Tip-enhanced quantum-sensing spectroscopy for hBN single-photon emitters.} 
(a) Schematic illustration of deterministic emitter-cavity coupling and tip-enhanced single-photon emission.
(b) Energy level diagrams of the defect-state transitions in hBN for an uncoupled single-photon emitter (left) and the same single-photon emitter when coupled to the tip-cavity (right).
The overlaid contour map (red) in the right panel highlights the calculated spatial distribution of the spontaneous emission rate enhancement in the tip-cavity.
(c) Top: Representative normalized photoluminescence spectra from multiple hBN single-photon emitters, each exhibiting a distinct ZPL. 
Bottom: Normalized tip-plasmon spectra, showing a tunable spectral overlap with the ZPL energies of the hBN single-photon emitters.
Abbreviations: single-photon emitter (SPE); photoluminescence intensity (\textit{I}$_\textnormal{PL}$); scattering intensity (\textit{I}$_\textnormal{scatt}$).
}
	\label{fig:fig1}
\end{figure*}

\noindent
\textbf{Spatio-spectrally deterministic emitter-cavity coupling}

\noindent
Our tip-enhanced quantum-sensing spectroscopy based on a shear-force atomic force microscope (AFM) system (Experimental methods and Fig. S1 in Supplementary Information) \cite{lee2024}.
This design enables 3-dimensional cavity positioning with sub-nanometer precision (Fig. 1a), ensuring high-fidelity coupling between atomic-scale emitters and the plasmonic cavity systems. 
As shown in Fig. 1b, an uncoupled hBN single-photon emitter exhibits the intrinsic spontaneous emission dynamics of the atomic defect. 
In contrast, when the same emitter is coupled to the plasmonic tip-cavity, its spontaneous emission rate is significantly enhanced by the Purcell effect \cite{purcell1946, akselrod2014}.
Notably, our tip geometry maximizes radiative decay pathways, efficiently converting the enhanced emission into detectable photons \cite{matsuzaki2021, rogobete2007, mohammadi2010}.

To optimize the coupling conditions, we begin by characterizing the spectral properties of both the single-photon emitters in hBN and the plasmons of tip-cavities (Fig. S2-3 in Supplementary Information).
Fig. 1c presents normalized photoluminescence spectra from several hBN single-photon emitters, each exhibiting a narrow ZPL at a distinct energy within the visible spectral range, set by the defect-specific electronic level separation \cite{tran2016}.
The corresponding plasmon resonance spectra demonstrate that the cavity resonance can be tuned to effectively overlap with the various ZPL energies (Fig. S4 in Supplementary Information).
By precisely matching the plasmon resonance to each target ZPL or excitation energy, we selectively control the excitation and emission rates for that emitter.
Consequently, this platform enables a comprehensive investigation of how light-matter interactions influence single-photon emission characteristics, providing a versatile framework for fundamental studies and future quantum photonic applications.
\\

\begin{figure*}[!t]
	\includegraphics[width = 14.0 cm]{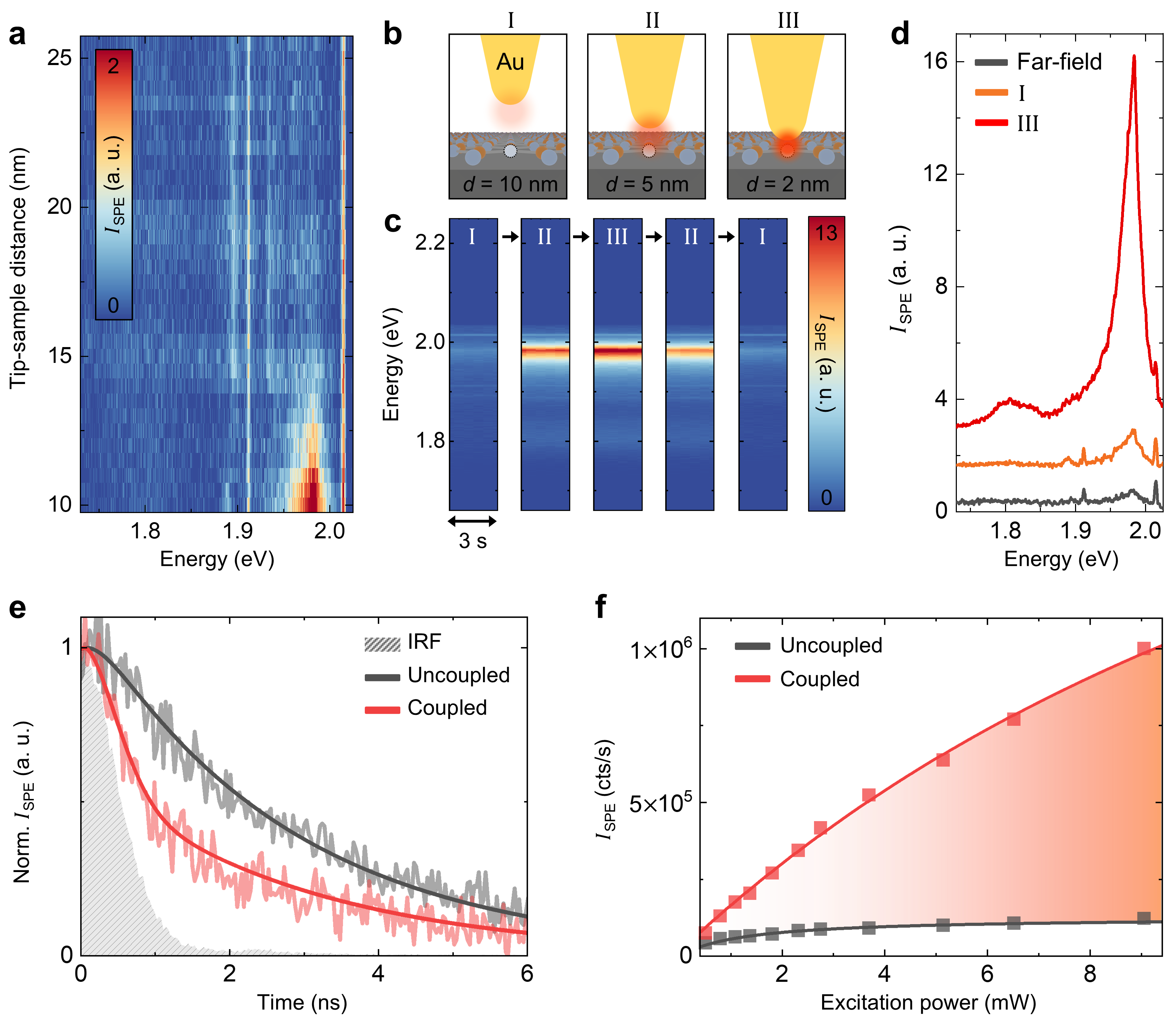}
	\caption{
\textbf{Distance-dependent spectroscopic sensing of hBN single-photon emitters.} 
(a) Evolution of the single-photon emission spectra as a function of tip–sample distance \textit{d}.
(b) Schematic of three regimes defined by \textit{d}-dependence.
(c) Systematic tuning of the single-photon emission intensity for regime I (\textit{d} = 10 nm), regime II (\textit{d} = 5 nm), and regime III (\textit{d} = 2 nm).
(d) Representative single-photon emission spectra showing the evolution of emission brightness from the far-field (black) to regime I (orange) and regime III (red).
(e) Normalized time-resolved single-photon emission decay traces of hBN single-photon emitter uncoupled (black) and coupled (red) to the plasmonic cavity.
(f) Single-photon emission intensity as a function of excitation power for an uncoupled hBN single-photon emitter (black) and the same single-photon emitter coupled to the plasmonic cavity (red).
}
	\label{fig:fig2}
\end{figure*}

\noindent
\textbf{Tip-cavity–enabled deterministic high-brightness hBN single-photon emitters}

\noindent
We investigate the distance-dependent coupling between the hBN single-photon emitter and the plasmonic tip-cavity by precisely regulating the van der Waals force between them.
As shown in Fig. 2a, the single-photon emission spectra exhibit pronounced changes as the plasmonic tip approaches the hBN single-photon emitter.
The emission intensity increases exponentially at distances of $<$15 nm, consistent with near-field coupling in plasmonic systems.
Since the thickness of the hBN flake, only the single-photon emitters located near its surface are enhanced by the Au tip, while the Raman signal from the Si substrate outside the near-field region remains unchanged. 
We attribute the observed enhancement to two main mechanisms: (i) increased excitation rates due to localized plasmon and electrostatic lightning rod effects \cite{lee2020, mohamed2000}, and (ii) the enhancement of spontaneous emission rates by the Purcell effect in the cavity \cite{purcell1946, akselrod2014}.
By using our 3-dimensional cavity alignment system featuring a switchable Au tip and/or emitter, we can systematically control the collective contributions of these two mechanisms.
Stepwise adjustments of the tip-sample distance (\textit{d}) show three regimes, each yielding a stable and reversible level of emission enhancement (Fig. 2b-c).
Comparative emission spectra (Fig. 2d) highlight the progressive increase in brightness from the far-field condition to regime I (\textit{d} = 10 nm) and then to regime III (\textit{d} = 2 nm), with the closest approach producing more than an order-of-magnitude enhancement in brightness.
This systematic control of coupling strength through precise spatial positioning represents a significant advance in the manipulation of quantum emitter-cavity interactions, enabling dynamic tuning of key photonic properties, such as brightness, emission lifetime, and photon statistics.
We demonstrate that coupling hBN single-photon emitters to the plasmonic tip-cavity dramatically increases their spontaneous emission rates through the Purcell effect, fundamentally overcoming the photon bottleneck that limits high-flux quantum emission under intense excitation conditions \cite{you2025}. 
To confirm plasmonic coupling and quantify the enhancement mechanism, we measure the excited-state lifetime of the hBN single-photon emitter through deconvolution of the instrument response function (IRF) using least-squares fitting \cite{hoang2015, pang2019}, as shown in Fig. 2f. 
Time-resolved photoluminescence decay traces reveal a pronounced lifetime reduction from $\sim$2.9 ns for the uncoupled emitter to $\sim$89.0 ps upon coupling, corresponding to the Purcell effect that directly accounts for the substantially elevated saturation power and enhanced photon output of the coupled system.
Fig. 2f shows a comparison of the single-photon emission intensity as a function of excitation power between an uncoupled hBN single-photon emitter and the same emitter coupled to the tip-cavity.
In the uncoupled system, the intrinsically limited spontaneous emission rate creates a fundamental photon bottleneck that causes emission saturation at moderate pump powers, preventing efficient utilization of the incident excitation and constraining the achievable photon flux. 
By contrast, the cavity-coupled single-photon emitter benefits from Purcell-enhanced spontaneous emission rates, enabling rapid excited-state depletion and sustained bright single-photon output over a extended excitation range.
Quantitative analysis using a first-order saturation model reveals saturation powers of $\sim$1.3 mW for the uncoupled emitter and $\sim$16.7 mW for the coupled emitter, demonstrating that plasmonic coupling sustains substantially higher brightness under strong excitation \cite{tran2016}. 
This enhancement translates to a remarkable increase in maximum single-photon emission rate from $\sim$1.3$\times$10$^5$ cts/s to $\sim$2.8$\times$10$^6$ cts/s, representing a $\sim$21.5-fold improvement in single-photon flux. 
This advantageous behavior is sustained at the highest excitation powers applied, where the cavity-coupled emitter maintains bright single-photon emission while effectively suppressing bunching-induced degradation of single-photon purity that typically compromises quantum emission at high pump rates \cite{tran2016, maragkou2012, babinec2010, kurtsiefer2000}. 
This observation reveals the potential of engineered plasmonic nanocavities to simultaneously optimize both photon flux and quantum purity, paving the way toward high-performance single-photon sources required for practical quantum technologies.
\\

\begin{figure*}[!t]
	\includegraphics[width = 13.5 cm]{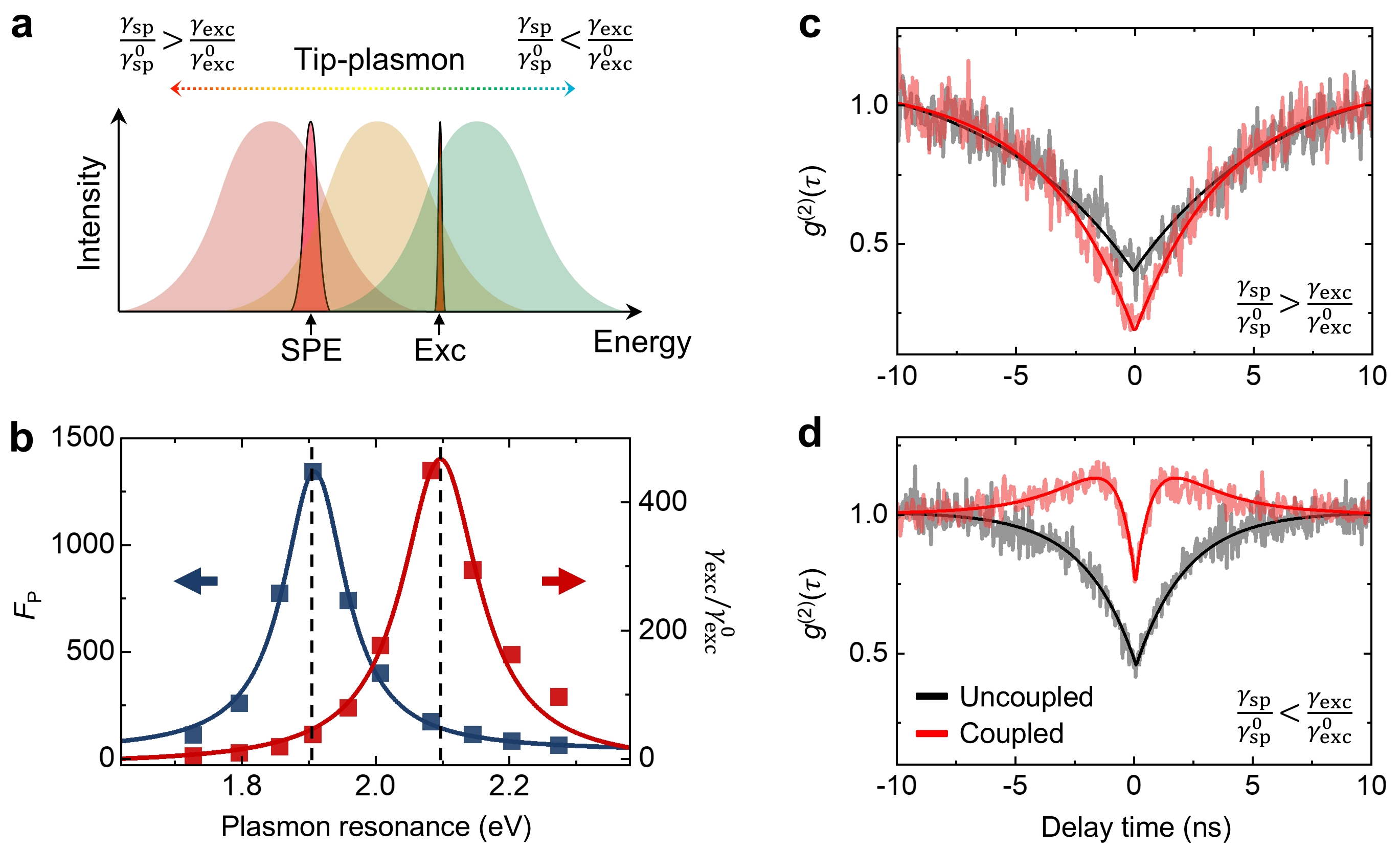}
	\caption{
\textbf{Reconfigurable brightness and single-photon purity of hBN single-photon emitters.} 
(a) Conceptual illustration of the interplay among the single-photon emission (SPE), the excitation laser (Exc), and the plasmon of tip-cavity.
By tuning the spectral overlap of these resonances (indicated by colored regions), either the excitation rate (${\gamma}_{\textnormal{exc}}^{}$/${\gamma}_{\textnormal{exc}}^{0}$) or spontaneous emission rate (${\gamma}_{\textnormal{sp}}^{}$/${\gamma}_{\textnormal{sp}}^{0}$) can be selectively enhanced. 
(b) Peak values of \textit{F}$_\textnormal{P}$ (blue, left axis) are evaluated at the emitter position (650 nm, $\sim$1.91 eV) and peak excitation-rate enhancement ${\gamma}_{\textnormal{exc}}^{}$/${\gamma}_{\textnormal{exc}}^{0}$ (red, right axis) are evaluated at the laser excitation wavelength (594 nm, $\sim$2.09 eV), both plotted against the plasmon energy.
Measured second-order correlation functions \textit{g}$^{(2)}$($\uptau$) of uncoupled (black) and cavity-coupled (red) emission under two coupling conditions: ${\gamma}_{\textnormal{sp}}^{}$/${\gamma}_{\textnormal{sp}}^{0}$$>$${\gamma}_{\textnormal{exc}}^{}$/${\gamma}_{\textnormal{exc}}^{0}$ (c) and ${\gamma}_{\textnormal{sp}}^{}$/${\gamma}_{\textnormal{sp}}^{0}$$<$${\gamma}_{\textnormal{exc}}^{}$/${\gamma}_{\textnormal{exc}}^{0}$ (d). 
}
	\label{fig:fig3}
\end{figure*}

\noindent
\textbf{Reconfigurable photon statistics in tip-enhanced quantum-sensing}

\noindent
Fig. 3a illustrates how the interplay among the energies of optical transition in the single-photon emitter, the energy of excitation laser, and the tip-plasmon of the cavity governs the single-photon emission behavior.
To maintain high single-photon purity, we deliberately avoid aligning electronic resonance of an emitter with the excitation energy, as fully resonant excitation would substantially degrade the single-photon purity.
Consequently, the detuning between the excitation energy and the ZPL energy allows us to independently tune the excitation rate versus the spontaneous emission rate via systematically controlled plasmonic coupling, a clear distinction from coupling approaches in conventional emitters. 
For example, when the plasmon resonance is red-shifted to strongly overlap with the emission resonance of the single-photon emitter while remaining detuned from the excitation energy, the spontaneous emission rate is preferentially enhanced over the excitation rate (${\gamma}_{\textnormal{sp}}^{}$/${\gamma}_{\textnormal{sp}}^{0}$$>$${\gamma}_{\textnormal{exc}}^{}$/${\gamma}_{\textnormal{exc}}^{0}$). 
In contrast, when it is blue-shifted toward the laser energy, excitation rate is enhanced more strongly than spontaneous emission rate (${\gamma}_{\textnormal{sp}}^{}$/${\gamma}_{\textnormal{sp}}^{0}$$<$${\gamma}_{\textnormal{exc}}^{}$/${\gamma}_{\textnormal{exc}}^{0}$).
We demonstrate selective control of excitation and spontaneous emission enhancements by spectrally positioning the tip-plasmon relative to the laser excitation and the ZPL of the hBN emitter. 
To quantify this, we perform 3-dimensional finite‐difference time‐domain (FDTD) simulations for a series of plasmon resonance spanning the experimentally observed range (1.7–2.3 eV). 
In each simulation, we compute (i) the normalized excitation‐rate enhancement (${\gamma}_{\textnormal{exc}}^{}$/${\gamma}_{\textnormal{exc}}^{0}$) by illuminating the plasmonic geometry with a plane wave at varying photon energies and recording the local intensity increase at the nominal emitter location, and (ii) Purcell factor (\textit{F}$_\textnormal{P}$) by embedding an oscillating electric dipole at the same position and evaluating its spontaneous emission rate enhancement. 
The resulting spectra reveal that, although both ${\gamma}_{\textnormal{exc}}^{}$/${\gamma}_{\textnormal{exc}}^{0}$ and \textit{F}$_\textnormal{P}$ derived from the same underlying plasmonic scattering cross‐section, by virtue of electromagnetic reciprocity \cite{bedingfield2022}, their peak overlaps with the excitation ($\sim$2.09 eV) and emission ($\sim$1.91 eV) energies differ in energy and amplitude (Fig. S5 in Supplementary Information).
To quantify these spectral trends and derive a concise design rule, we extract the absolute maxima of each enhancement curve and plot them versus the plasmon peak energy in Fig. 3d. 
Two distinct regimes become evident. For plasmon energies below $\sim$2.00 eV, the Purcell factor enhancement dominates, indicating that that the increase in spontaneous emission rate is the primary enhancement mechanism.  
On the other hand, for plasmon energies above $\sim$2.00 eV, excitation rate enhancement prevails, signifying a regime of preferential field-driven excitation. Notably, at the crossover near 2.00 eV, both processes are comparably enhanced, indicating a balanced regime where the modified excitation and emission contribute comparably to the overall enhancement.

These simulation results provide a direct, quantitative foundation for our observations in Fig. 3c–d. 
By selecting plasmonic resonances that lie below, at, or above the crossover energy, we deterministically tune the ratio ${\gamma}_{\textnormal{sp}}^{}$/${\gamma}_{\textnormal{sp}}^{0}$ to ${\gamma}_{\textnormal{exc}}^{}$/${\gamma}_{\textnormal{exc}}^{0}$. 
The Au tips with low‐energy plasmons yield a high \textit{F}$_\textnormal{P}$/${\gamma}_{\textnormal{exc}}^{}$ ratio, suppressing bunching and improving single-photon purity, whereas the tips with high‐energy plasmon minimize \textit{F}$_\textnormal{P}$/${\gamma}_{\textnormal{exc}}^{}$, boosting an intense pumping. The intermediate regime affords a tunable platform in which modest adjustments to tip geometry, or the local dielectric environment enable a smooth transition between these two extremes. 
As shown in Fig. 3c-d, these two contrasting coupling scenarios lead to markedly different photon statistics in the emitted light.
In the regime of ${\gamma}_{\textnormal{sp}}^{}$/${\gamma}_{\textnormal{sp}}^{0}$$>$${\gamma}_{\textnormal{exc}}^{}$/${\gamma}_{\textnormal{exc}}^{0}$, the rapid depletion of the excited state suppresses bunching events.
This leads to an improvement in single-photon purity, as reflected in a pronounced reduction of the \textit{g}$^{(2)}$(0) value from $\sim$0.40 to $\sim$0.17 upon cavity coupling.
In contrast, in the case of ${\gamma}_{\textnormal{sp}}^{}$/${\gamma}_{\textnormal{sp}}^{0}$$<$${\gamma}_{\textnormal{exc}}^{}$/${\gamma}_{\textnormal{exc}}^{0}$, the elevated bunching probability leads to degraded single-photon purity (Fig. S6-7 in Supplementary Information).
This manifests as a transition from sub-Poissonian photon statistics to pronounced photon bunching, evidenced by an increased \textit{g}$^{(2)}$(0) value from $\sim$0.46 to $\sim$0.74.
Beyond validating the controlled switching between anti-bunched and bunched photon statistics in our experiments, this framework establishes general design rules for cavity-coupled single-photon emitters. 
\\

\noindent
\textbf{Tip-enhanced quantum-sensing of single-defect magnetic resonance in hBN}

\noindent
For tip-enhanced quantum-sensing, we can directly attach an hBN nanoflake at the tip apex (Fig. 4a-c), yielding bright, localized single-photon emission confined to the apex region (Fig. S8 in Supplementary Information).
The geometric asymmetry of the tip-based plasmonic nanocavity produces a markedly polarization-dependent near-field distribution. 
FDTD calculations (Fig. 4d) show that p-polarized excitation (parallel to the tip axis) induces field localization at the emitter site via the combined action of the tip-plasmon and the lightning-rod effect, whereas s-polarized excitation generates a substantially weaker field. 
This polarization selectivity directly appears in the single-photon spectra (Fig. 4e): under p-polarization the emitter experiences a higher effective excitation rate and emits more intensely than under s-polarization.

\begin{figure*}
	\includegraphics[width = 16.0 cm]{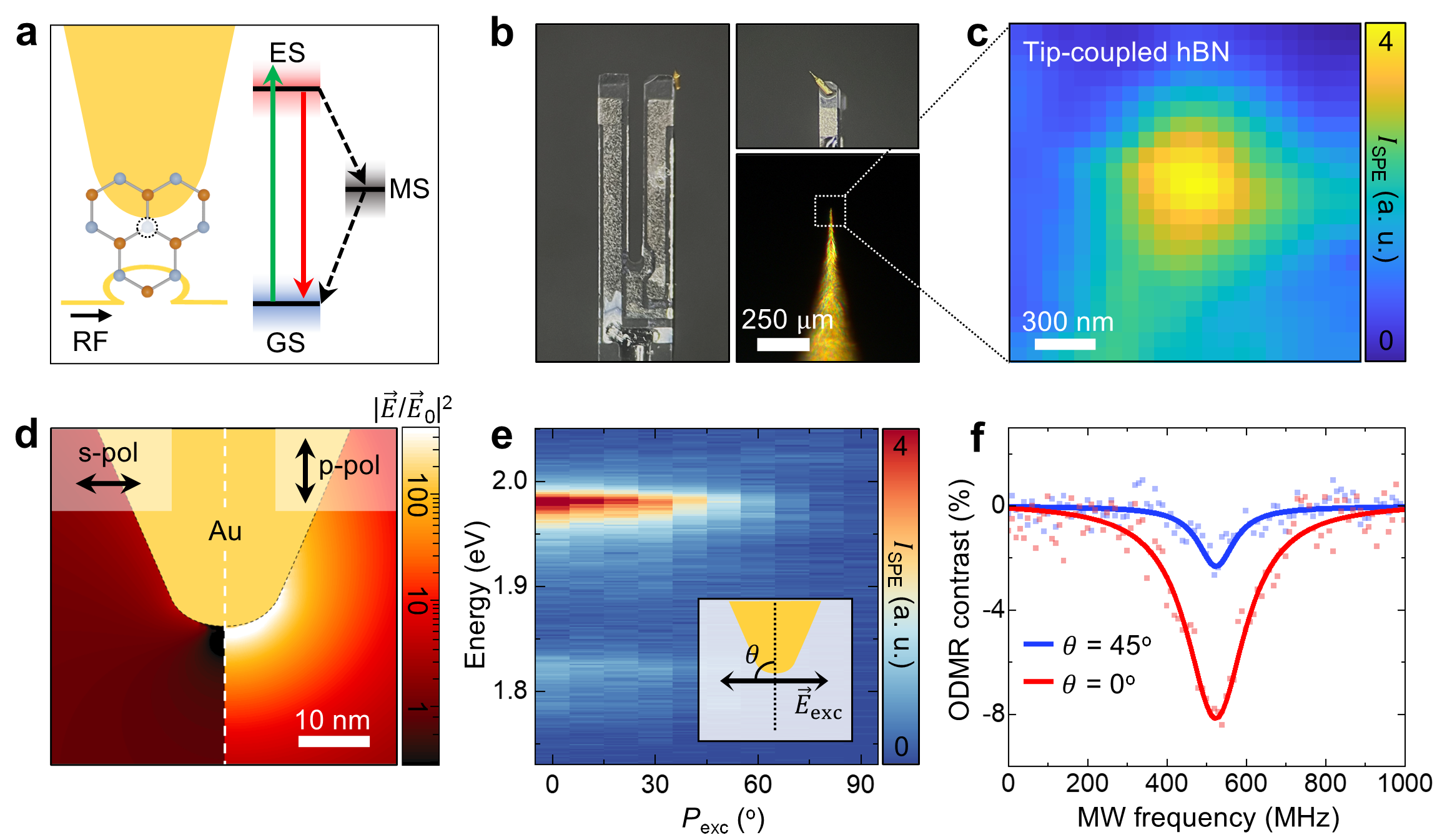}
	\caption{
\textbf{Optically detected magnetic resonance of a hBN single defect integrated to the plasmonic tip.}
(a) Schematic of tip-enhanced quantum-sensing with a hBN single defect on the Au tip.
(b) Microscope images of the plasmonic tip mounted on the quartz tuning fork resonator. 
(c) SPE intensity map of a single defect embedded within the hBN nanoflake at the tip apex.
(d) Simulated plasmonic field distributions under s-polarized (left) and p-polarized (right) excitation, showing distinct field enhancements in the plasmonic nanocavity.
(e) Single-photon emission spectra as a function of excitation polarization angle (\textit{P}$_\textnormal{exc}$). 
An excitation polarization of 0$^\circ$ corresponds to alignment parallel to the tip axis, while 90$^\circ$ corresponds to alignment perpendicular to the tip axis.
(f) ODMR spectra of a single defect at the tip apex under 45$^\circ$ (blue) and 0$^\circ$ (red) excitation.
}
	\label{fig:fig4}
\end{figure*}

To assess plasmonic-coupling effects on spin readout, we recorded ODMR spectra of the single defect at the tip apex for different excitation polarizations (Fig. 4f). 
For 45$^\circ$ excitation, we obtain the ODMR contrast of $\sim$2.3$\%$, the linewidth of $\sim$110 MHz, and the photon detection rate of $\sim$188 kcts/s, which yield a shot-noise-limited DC magnetic-field sensitivity $\eta$ of $\sim$303 $\mu$\textit{T}$/$$\sqrt{\textnormal{Hz}}$ using
\begin{equation} 
	\begin{array}{cl}
		\eta = A\times\cfrac{h}{g\mu_\textnormal{B}}\times\cfrac{\Delta\nu}{C\sqrt{R}}
	\end{array}
	\label{eq1}
\end{equation}
\\
, where \textit{A} is a numerical paramter related to the line shape profile of the resonance peak, \textit{h} is the Planck constant, \textit{g} is the Lande factor, $\mu$$_\textnormal{B}$ is the Bohr magneton, $\Delta$$\nu$ is the ODMR linewidth, \textit{C} is the ODMR contrast, \textit{R} is the photon count rate.
In contrast, for 0$^\circ$ (p-polarized) excitation, the full exploitation of the plasmon-induced localized fields leads to the enhanced ODMR contrast of $\sim$8.3$\%$ and photon detection rate of $\sim$263 kcts/s, improving the sensitivity to $\sim$116 $\mu$\textit{T}$/$$\sqrt{\textnormal{Hz}}$.
In the low-excitation regime, where optical pumping rates are lower than the decay rates, the increase in radiative decay rates reduces the ODMR contrast.
This reduction arises from a shortened radiative lifetime due to an increased photonic density of states, which disproportionately lowers the relative probability of nonradiative decay pathways \cite{bogdanov2017}. 
These pathways are more prevalent in the spin sublevels with higher intersystem crossing rates, thereby diminishing the differential emission between spin states.
In the optical saturation regime, characterized by high excitation rates that maintain a continuously populated excited state, the ODMR contrast becomes independent of the radiative lifetime.
Instead, it is governed solely by nonradiative transition rates, such as intersystem crossing and singlet decay rates \cite{bogdanov2017}.
As a result, the Purcell effect significantly enhances the spin readout signal-to-noise ratio (SNR) by increasing photon flux, without compromising ODMR contrast, as spin selectivity remains determined by these nonradiative processes.
These findings establish tip-integrated hBN single-photon emitters as a promising platform for nanoscale quantum magnetometry. 
By engineering light-matter interactions, we improve magnetic-field sensitivity, while vdW material-embedded architectures offer surface-proximal and 2D-compatible operation that can complement single nitrogen vacancy (NV)-center probes, potentially unlocking unique application pathways in quantum sensing.
\\

\noindent
{\bf Discussion}
\\
\noindent
We have introduced tip-enhanced quantum-sensing spectroscopy that deterministically couples plasmonic tip-cavities to atom-like defects in hBN and reconfigures their quantum-optical response. 
By spatio-spectrally positioning the cavity relative to the emitter transition, the excitation laser, and the plasmon resonance, we selectively enhance the excitation and/or spontaneous-emission channels. 
This control removes the photon-bottleneck imposed by intrinsically slow radiative decay, enabling sustained high-flux operation while preserving single-photon purity. 
Consistent with our measurements and FDTD-derived design rule, plasmon energies below the crossover preferentially increase the spontaneous emission rates, which shortens the excited-state lifetime and suppresses bunching, whereas blue-shifted plasmons predominantly enhance the excitation rate, promoting bunching and reducing purity. 
The ability to dial between these regimes provides a practical route to reconfigurable brightness and photon statistics in solid-state single-photon sources.
Beyond spectral alignment, non-quasi-static, geometrically asymmetric plasmonic nanocavities furnish an additional lever: under plane-wave drive they couple efficiently to odd multipolar modes, such that the radiative decay rate can exceed the excitation-rate enhancement without violating electromagnetic reciprocity \cite{bedingfield2022, kongsuwan2020}. 
This asymmetric coupling offers a general pathway to on-demand decoupling and rebalancing of excitation and emission rates in cavity-integrated emitters.
The same control principle benefits quantum sensing. 
By operating in the optical-saturation regime, Purcell-accelerated emission maximizes photon flux without degrading spin selectivity, thereby improving shot-noise-limited ODMR sensitivity while maintaining contrast determined by nonradiative pathways. 
Together, these results establish a versatile, deterministic platform for manipulating quantum emission characteristics across diverse solid-state systems and for tip-integrated nanoscale magnetometry, bridging fundamental cavity QED with the high-rate, high-fidelity requirements of emerging quantum technologies.
\\

\noindent
\textbf{Data availability.}
The data that support the plots within this paper and other findings of this study are available from the corresponding author upon reasonable request.

\bibliography{hBN} 

\vskip 1cm
\noindent
{\bf Acknowledgements}

\noindent
This work was supported by the National Research Foundation of Korea (NRF) grants (RS-2024-00440375).

\noindent
{\bf Author contributions}

\noindent
H.L., T.M. contributed equally to this work.
H.L., T.M., M.T., I.A., K.-D.P. conceived the experiments.
H.L., T.M. performed all experimental measurements.
H.L., T.M., H.O. optimized and prepared the hBN single-photon emitter samples.
H.L., H.J., K.P. performed the theoretical calculations.
H.L., T.M., M.T, I.A., K.-D.P. analyzed the data, and all authors discussed the results.
H.L., T.M., K.-D.P. wrote the manuscript with contributions from all authors.
K.-D.P. supervised the project.

\noindent
{\bf Additional information}

\noindent
Supplementary information is available for this paper.

\noindent
{\bf Competing financial interests}

\noindent
The authors declare no competing financial interests.

\end{document}